\documentclass[aps,prd,preprint,superscriptaddress,tightenlines,nofootinbib]{revtex4}

\long\def\symbolfootnote[#1]#2{\begingroup%
\def\thefootnote{\fnsymbol{footnote}}\footnote[#1]{#2}\endgroup}

\usepackage{amsfonts,amsmath,amsthm,amssymb}
\usepackage{mathrsfs}
\usepackage{bm}
\usepackage{color}
\usepackage{epsfig}

\newcommand{\PRE}[1]{{#1}}   

\def\thalf{\tfrac{1}{2}}
\def\mpl{M_{\rm Pl}}

\newcommand{\douba}[2]{\ensuremath{                                                  
\left( \begin{array}{c} #1    \\ #2 
        \end{array}\right)}}

\newcommand{\beq}{\begin{equation}}
\newcommand{\eeq}{\end{equation}}
\newcommand{\bea}{\begin{flushleft} \begin{eqnarray}}
\newcommand{\eea}{\end{eqnarray}\end{flushleft}}

\newcommand{\postscript}[2]{\setlength{\epsfxsize}{#2\hsize}
   \centerline{\epsfbox{#1}}}
\newcommand{\comment}[1]{}

\newcommand{\ci}[1]{}

\newcommand{\lb}{\left(}
\newcommand{\rb}{\right)}

\newcommand{\ba}{\begin{eqnarray}}
\newcommand{\ea}{\end{eqnarray}}
\newcommand{\be}{\begin{equation}}
\newcommand{\ee}{\end{equation}}
\newcommand{\bay}[1]{\left(\begin{array}{#1}}
\newcommand{\eay}{\end{array}\right)}

\def\xe{{\epsilon}}

\definecolor{rossoCP3}{cmyk}{0,.88,.77,.40}

\begin{document}


%

\title{\color{rossoCP3}{\PRE{\vspace*{0.9in}}
Reconciling BICEP2 and Planck results with right-handed Dirac 
neutrinos in the fundamental representation of  grand unified $\bm{E_6}$ 
\PRE{\vspace*{0.1in}} }}

\author{ Luis A. Anchordoqui}
\affiliation{Department of Physics,\\
University of Wisconsin-Milwaukee,
 Milwaukee, WI 53201, USA
\PRE{\vspace*{.05in}}
}

\author{Haim \nolinebreak Goldberg}
\affiliation{Department of Physics,\\
Northeastern University, Boston, MA 02115, USA
\PRE{\vspace*{.05in}}
}

\author{Xing Huang}
\affiliation{Department of Physics, \\
National Taiwan Normal University, Taipei, 116, Taiwan
\PRE{\vspace*{.05in}}
}

\author{Brian J. Vlcek}
\affiliation{Department of Physics,\\
University of Wisconsin-Milwaukee,
 Milwaukee, WI 53201, USA
\PRE{\vspace*{.05in}}
}

\date{April 2014}

\PRE{\vspace*{.15in}}

\begin{abstract}
\PRE{\vspace*{.15in}}
\noindent 
  The tensor-to-scalar ratio ($r = 0.20^{+0.07}_{-0.05}$) inferred
  from the excess B-mode power observed by the Background Imaging of
  Cosmic Extragalactic Polarization (BICEP2) experiment is almost
  twice as large as the 95\% CL upper limits derived from temperature
  measurements of the WMAP ($r<0.13$) and Planck ($r<0.11$) space
  missions. Very recently, it was suggested that additional
  relativistic degrees of freedom beyond the three active neutrinos
  and photons can help to relieve this tension: the data favor an
  effective number of light neutrino species $N_{\rm eff} = 3.86 \pm
  0.25$.  Since the BICEP2 ratio implies the energy scale of inflation
  ($V_*^{1/4} \sim 2 \times 10^{16}~{\rm GeV}$) is comparable to the
  grand unification scale, in this paper we investigate whether we can
  accommodate the  required $N_{\rm eff}$ with three right-handed
  (partners of the left-handed standard model) neutrinos living in the
  fundamental representation of a grand unified exceptional $E_6$
  group. We show that the superweak interactions of these Dirac states
  (through their coupling to a TeV-scale $Z'$ gauge boson) lead to
  decoupling of right-handed neutrino just above the QCD cross over
  transition: $175~{\rm MeV} \alt T_{\nu_R}^{\rm dec} \alt 250~{\rm MeV}$. For
  decoupling in this transition region, the contribution of the three
  right-handed neutrinos to $N_{\rm eff}$ is suppressed by heating of
  the left-handed neutrinos (and photons). Consistency (within
  $1\sigma$) with the favored $N_{\rm eff}$ is achieved for $4.5~{\rm TeV} < M_{Z'} < 7.5~{\rm TeV}$. The
  model is fully predictive and can be confronted with future data
  from LHC14.
\end{abstract}

\maketitle

\section{Introduction}

The concordance model of cosmology, with dark energy ($\Lambda$), cold
dark matter (CDM), baryons, and three flavors of left-handed ({\it
  i.e.} one helicity state $\nu_L$) neutrinos (along with their
right-handed antineutrinos $\overline \nu_R$), provides a consistent
description of the late early universe: big-bang nucleosynthesis
(BBN), at $\sim 20$ minutes, the cosmic microwave background (CMB), at
$\sim 380~{\rm Kyr}$, and the galaxy formation epoch, at $\agt 1~{\rm
  Gyr}$~\cite{Beringer:1900zz}.  Inflationary cosmology extends the
$\Lambda$CDM model by postulating an early period where the scale
factor of the universe expands exponentially: $a \propto e^{Ht}$,
where $H = \dot a/a$ is the Hubble parameter~\cite{Guth:1980zm}. If
the interval of exponential expansion satisfies $\Delta t \agt
60/H$, a small casually connected region can grow sufficiently to
accommodate the observed homogeneity and isotropy, to dilute any
overdensity of magnetic monopoles, and to flatten the spatial
hyper-surfaces (i.e., $\Omega \equiv \frac{8\pi \rho}{3 M_{\rm Pl} H^2}
\to 1$, where $M_{\rm Pl} = G^{-1/2}$ is the Planck mass and $\rho$
the energy density; throughout we use natural units, $c = \hslash = 1$).

The simplest inflationary models adopt Einstein gravity sourced by a
scalar field $\phi$ and a potential
$V(\phi)$~\cite{Linde:1981mu,Albrecht:1982wi,Linde:1983gd,Freese:1990rb}. In
co-moving coordinates an homogeneous scalar field with minimal coupling
to gravity has the equation of motion
\begin{equation} 
\ddot{\phi} + 3 H\dot{\phi} + V' = 0, 
\end{equation}
where $V' = dV/d\phi$. The phase of
quasi-de Sitter expansion ($H \approx {\rm const.}$), when the scalar field rolls slowly down
the potential, can only be sustained for a sufficient long period of
time if 
\be
\thalf \dot{\phi}^2 \ll  |V|  
\quad
{\rm and} \quad 
\left| \frac{\ddot{\phi}} {3 H \dot{\phi}}\right| \ll  1 \, .  
\ee
These conditions imply \be
\epsilon \equiv  \frac{\mpl^2}{16\pi} \left(\frac{ V'}{V}\right)^2\ll 1
\quad {\rm and} \quad 
\eta \equiv  \frac{\mpl^2} {8\pi} \left| \frac{ V''} {V} -
  \frac{1}{2} \left(\frac{V'}{V} \right)^2\right| \ll 1 \,,\ee
respectively.  

Quantum fluctuations in de Sitter space causally
generate large-scale density fluctuations, which are necessary for the
formation of galaxies and large-scale structure. As a bonus, small
perturbations ($h_{ij}$, with $h_i^i = \partial^i h_{ij} =0$) in the metric of space-time,
\begin{equation}
ds^2 \equiv g_{\mu \nu} dx^\mu \, dx^\nu = dt^2 - a \, (\delta_{ij} + h_{ij}) \, dx^i dx^j \,,
\end{equation}
become redshifted out to the
horizon~\cite{Mukhanov:1981xt,Hawking:1982cz,Bardeen:1983qw}. The
gravity-wave fluctuations are nearly frozen on super-Hubble scales and
their B-mode power spectrum,
\begin{eqnarray}
{\cal P}_h  & = &  A_t \lb \frac k {k_*} \rb^{n_t + \frac 1 2 \alpha_t
  \ln \lb \frac k {k_*} \rb + \cdots} \\ & \simeq &    {128V\over 3M_{\rm{Pl}}^4}\left[1 - \left(2C + \frac 5
    3 \right) \epsilon \right] \, \lb \frac k {k_*} \rb^{n_t + \frac 1 2 \alpha_t
  \ln \lb \frac k {k_*} \rb + \cdots} , 
\end{eqnarray} 
can be imprinted in the CMB temperature and
polarization. Here, the pivot $k_* = a H$ typifies scales probed by the CMB, $C \equiv \gamma_E + \ln 2 -2 \approx -0.7296$. To second order in $\epsilon$ the spectral index and its running are given by
\begin{equation}
n_t \simeq  -2\epsilon + \left(\frac 8 3 +4C \right) \xe \eta -\frac 2 3 (7+6C) \xe^2 
\quad
{\rm and} \quad
\alpha_t \equiv  {d n_t \over d \ln k}  \simeq  -4\epsilon(\epsilon-\eta)\,, \end{equation} 
respectively~\cite{Leach:2002ar}. On the other hand, the power spectrum of curvature perturbations is given by 
\begin{eqnarray}
{\cal  P}_\chi & = & A_s \lb \frac k {k_*} \rb^{n_s -1 + \frac 1 2
  \alpha_s  \ln \lb \frac k {k_*} \rb  + \cdots} \nonumber \\
& \simeq &  
 {8V\over 3M_{\rm{Pl}}^4\epsilon}\left[1 - (4C + 1)
    \epsilon + \lb 2C -\frac 2 3\rb\eta\right] \lb \frac k {k_*} \rb^{n_s -1 + \frac 1 2
  \alpha_s  \ln \lb \frac k {k_*} \rb  + \cdots} \,,
\end{eqnarray}
where \be n_s \simeq 1-4\epsilon+2\eta+ \left(\frac {10} 3 +4C \right)
\xe \eta - (6+4C) \xe^2 + \frac 2 3 \eta^2 - \frac{2}{3} (3 C-1)
\left(2 \xe^2-6 \xe \eta +\xi ^2\right)\,, \ee \be \alpha_s \equiv {d
  n_s \over d \ln k} \simeq -8 \epsilon ^2+ 16 \epsilon \eta
-2\xi^2\,, \quad {\rm and} \quad \xi^2 \equiv \frac {M_{\rm Pl}^4 V'
  V'''} {64 \pi^2 V^2} \, .  \ee 
For single field inflation with canonical kinetic term, the tensor spectrum shape is not independent from the other parameters. Slow-roll expansion implies a tensor-to-scalar ratio at the pivot scale of
\begin{equation}
 r\equiv {A_t \over A_s} \simeq 16 \epsilon + 32\lb C - \frac 1 3 \rb \xe (\xe - \eta) \, .
\end{equation}

Very recently, the BICEP2 Collaboration reported the measurement of
low-multipole B-mode polarization~\cite{Ade:2014xna}. The
observed B-mode power spectrum is well-fit by a $\Lambda$CDM $+ r$ model, 
with $r = 0.20^{+0.07}_{-0.05}$,
and is inconsistent with the null hypothesis, $r=0$, at a significance
of $7\sigma$. Such unexpectedly large value of $r$ corresponds to a
Hubble rate, $H \simeq 1.1 \times 10^{14}~{\rm GeV}$, constraining
the energy scale of inflation: $V_*^{1/4} \sim 2 \times 10^{16}~{\rm
  GeV}$. The BICEP2 dataset then provides the first experimental
evidence for the existence of a new physics scale in between the
electroweak and Planck scales, which is astonishingly closed to the
grand unification scale (determined by extrapolation of the running
coupling constants $\alpha_{\rm QCD}$, $\alpha_{\rm QED}$, and
$\alpha_{\rm weak}$ to a common, ``unified'' value).

BICEP2 data, however, is in significant tension with Planck's 95\% CL
upper limit, \mbox{$r< 0.11$,} from the temperature anisotropy
spectrum in the simplest inflationary $\Lambda$CDM~$+r$
model~\cite{Ade:2013uln}.\footnote{BICEP2 data is also in tension with
  the 95\% CL upper limit, \mbox{$r< 0.13$}, reported by the WMAP
  Collaboration~\cite{Hinshaw:2012aka}.}  The conflict is a result of
the fact that the large angle temperature excess foreshadowed by the
gravitational waves is not observed.  This apparent mismatch cannot be
resolved by varying parameters in this very restrictive, seven
parameter model: $\{ \Omega_{\rm CDM} h^2,$ $\Omega_b h^2,$ $\tau, \,
\Theta_{\rm s},\, A_s \,, n_s,\, r\}$, where $\Omega_{\rm CDM} h^2$ is
the CDM energy density, $\Omega_b h^2$ is the baryon density,
$\Theta_{\rm s}$ is the ratio between the sound horizon and the
angular diameter distance at decoupling, and $\tau$ is the Thomson
scattering optical depth of reionized intergalactic medium.

Several explanations have been put forward to help reconcile Planck
and BICEP2 measurements (see {\it e.g.}~\cite{Ashoorioon:2014nta,Ko:2014bka,Smith:2014kka}) . Of particular interest here, it
was pointed out that the tension can be relaxed if extra light species
({\it e.g.}  massive sterile neutrinos) contribute to the effective
number of relativistic degrees of freedom
(r.d.o.f.)~\cite{Giusarma:2014zza,Zhang:2014dxk,Dvorkin:2014lea}.  In
this work we take a somewhat related approach to investigate the
possibility of relaxing the tension by considering extra massless
neutrino species. Specifically, we associate the extra r.d.o.f. with
the right-handed partners of three Dirac neutrinos, which interact
with all fermions through the exchange of a new heavy vector meson
$Z'$.

\section{Constraints on Cosmological Parameters from CMB data}

As the BICEP2 Collaboration carefully emphasized~\cite{Ade:2014xna},
the measurement of $r = 0.2^{+0.07}_{-0.05}$ (or $r =
0.16^{+0.06}_{-0.05}$ after foreground subtraction, with $r=0$
disfavored at $5.9\sigma$) from the B-mode polarization appears to be
in tension with the 95\% CL upper limits reported by the WMAP ($r <
0.13$) and Planck ($r< 0.11$) collaborations from the large-scale CMB
temperature power spectrum. These upper limits, which favor
inflationary models with concave ($V''<0$) plateau-like inflaton
potentials, were derived on the basis of $\alpha_s =0$.

The discrepancy can  be formulated in terms of the tilt of $r$
\begin{equation}
{\cal T}_r = \frac{d \ln {\cal P}_h}{d\ln k} - \frac{d \ln {\cal P}_{\chi}}{d\ln k} = n_t - (n_s -1). 
\end{equation}
Planck data favor $n_s = 0.9603 \pm 0.0073$~\cite{Ade:2013uln}, whereas slow-roll inflationary
models yield the ``consistency relation,'' $n_t = -r/8$~\cite{Starobinsky:1985ww}, which is a
red-tilt for gravity waves. Consequently, for such inflationary
models, $n_t$ is negative and of order ${\cal O} (10^{-2})$. However,
BICEP2 and Planck reconciliation requires ${\cal T}_r \geq 0.16$,
which clearly shows the tension between standard slow-roll models with
Planck+BICEP2 data~\cite{Ashoorioon:2014nta}.

\begin{figure}[tbp]
\begin{minipage}[t]{0.49\textwidth}
\postscript{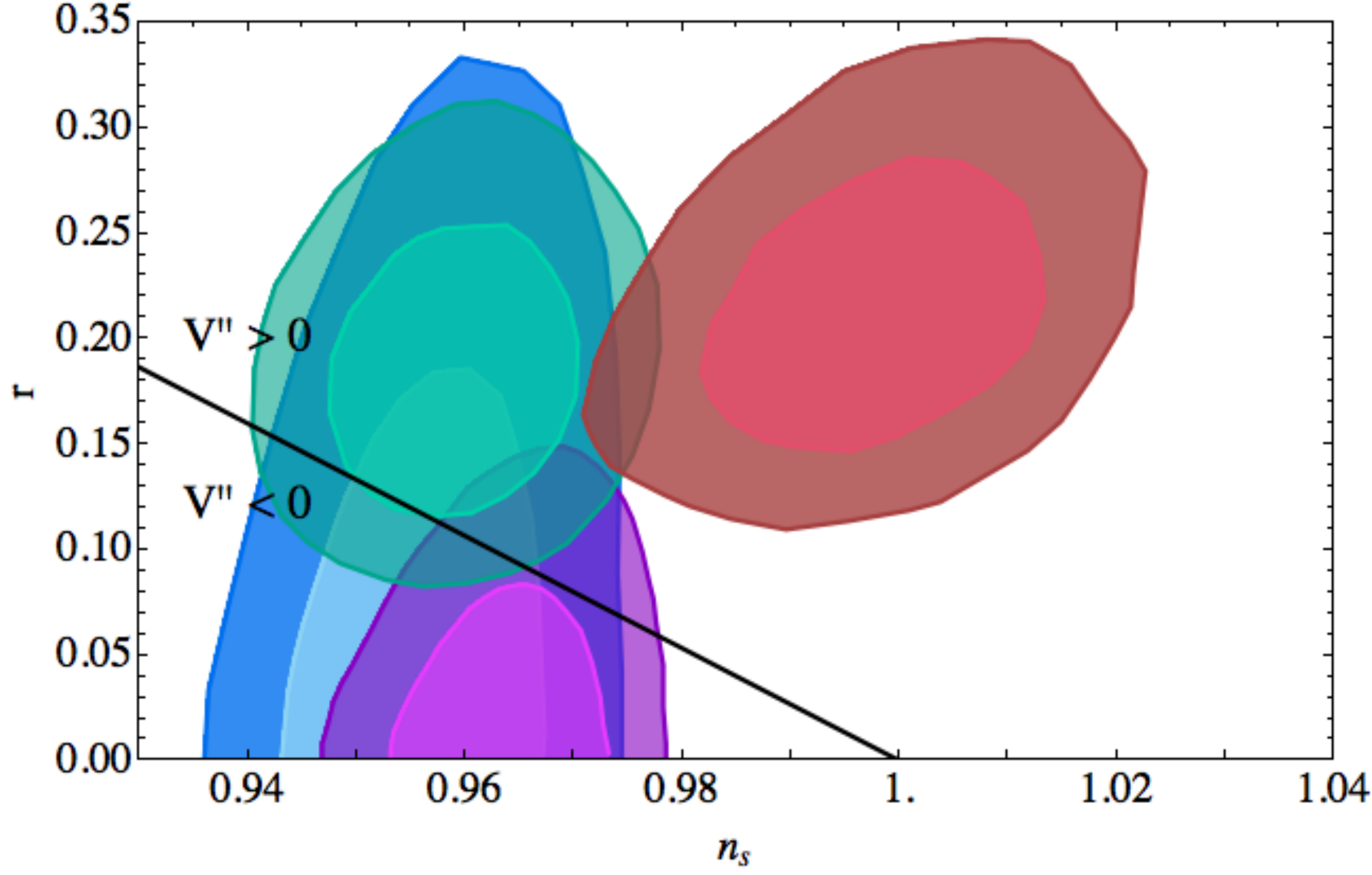}{0.99}
\end{minipage}
\hfill
\begin{minipage}[t]{0.49\textwidth}
\postscript{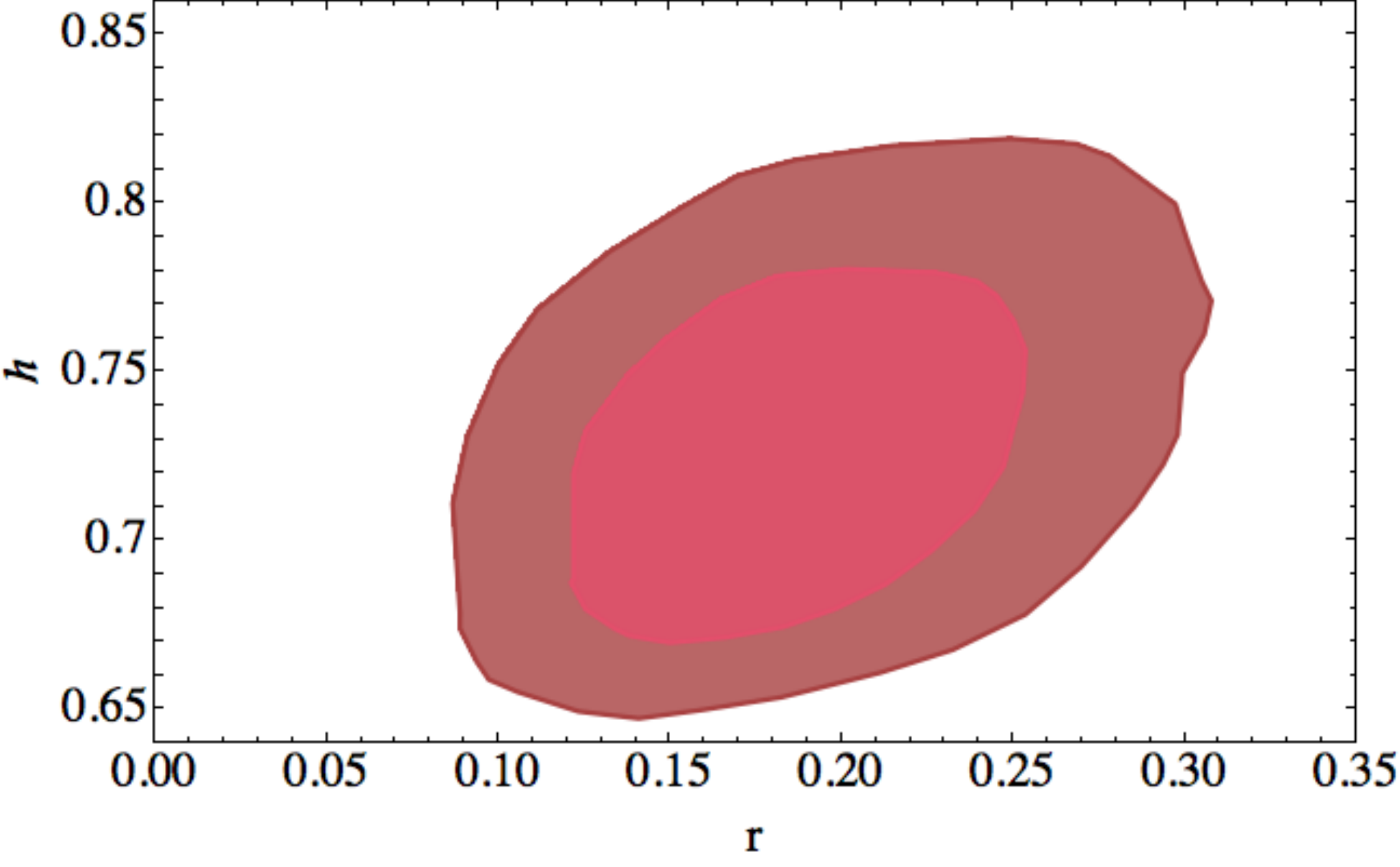}{0.99}
\end{minipage}
\caption{{\bf Left.} Marginalized joint 68\% CL and 95\% CL regions
  for ($r,n_s$) using Planck + WMAP + BAO with (blue) and without
  (purple) a running spectral index. The green contours show the
  BICEP2 $68\%$ and $95\%$ confidence regions for $(r,n_s)$, with
  $\alpha_s \neq 0$. The burgundy areas indicate the 68\% and 95\% CL
  allowed regions from the 9-parameter fit. The solid line shows the
  expected relation between $r$ and $n_s$, for a $V(\phi) \propto \phi$
  inflationary potential. {\bf Right.} 68\% and 95\% CL contours for
  ($h,r$).  The 9-parameter fit predicts a value of $h$ in concordance with
  observations by Planck and HST.}
\label{fig:uno}
\end{figure}

As shown in Fig.~\ref{fig:uno}, extension of the 7-parameter model to
include non-zero running of the spectral index ameliorates the
tension. However, the combination of Planck and BICEP2 data favors
$\alpha_s<0$ at almost the $3\sigma$ level, with best fit value around
$\alpha_s = -0.028 \pm 0.009~(68\% {\rm CL})$~\cite{Ade:2014xna}. This
is about 100 times larger than single-field inflation would
predict~\cite{Anchordoqui:2014uua}. Such a particular running can be
accommodated, however, if $V'''/V$ is roughly 100 times larger than
the natural expectation from the size of $V'/V \sim (10 M_{\rm
  Pl})^{-1}$ and $V''/V \sim (10 M_{\rm Pl})^{-2}$~\cite{Smith:2014kka}.

We previously noted in the Introduction that a higher effective number
of relativistic species can relieve the tension. To accommodate new
physics in the form of extra r.d.o.f., it is convenient to account for
the extra contribution to the standard model (SM) energy density, by
normalizing it to that of an ``equivalent'' neutrino species. The
number of ``equivalent'' light neutrino species,
\begin{equation}
N_{\rm eff} \equiv \frac{\rho_{\rm R} - \rho_\gamma}{\rho_{\nu_L}} \,,
\end{equation}
quantifies the total ``dark'' relativistic energy density (including
the three left-handed SM neutrinos) in units of the density of a
single Weyl neutrino: 
\begin{equation}
\rho_{\nu_L} = \frac{7 \pi^2}{120} \, \left(\frac{4}{11} \right)^{4/3}
T_\gamma^4,
\end{equation}
 where $\rho_\gamma$ is the energy density of photons
(with temperature $T_\gamma$) and $\rho_R$ is the total energy density
in relativistic particles~\cite{Steigman:1977kc}. Any relativistic
degree of freedom originating from physics beyond SM is included in
$N_{\rm eff}$. Allowing for the fact that the SM neutrinos do not decouple
instantaneously, which enables them to share some of the energy
released by $e^\pm$ annihilations, leads a small increase of the
contribution from the SM neutrino flavors to the number of
``equivalent'' light neutrino species, $N_{\rm eff} =
3.046$~\cite{Mangano:2005cc}.

\begin{figure}[tbp]
\begin{minipage}[t]{0.49\textwidth}
\postscript{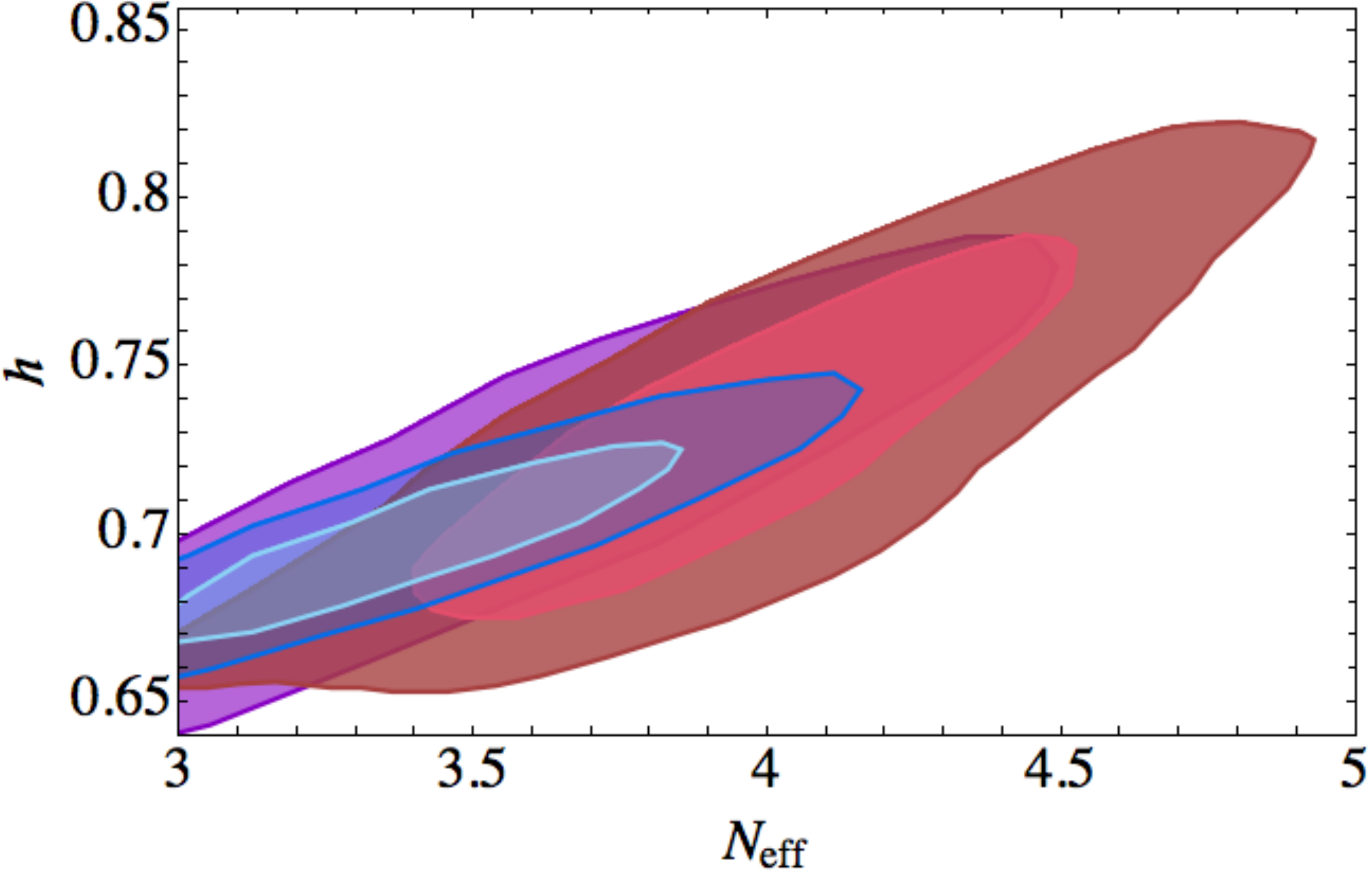}{0.99}
\end{minipage}
\hfill
\begin{minipage}[t]{0.49\textwidth}
\postscript{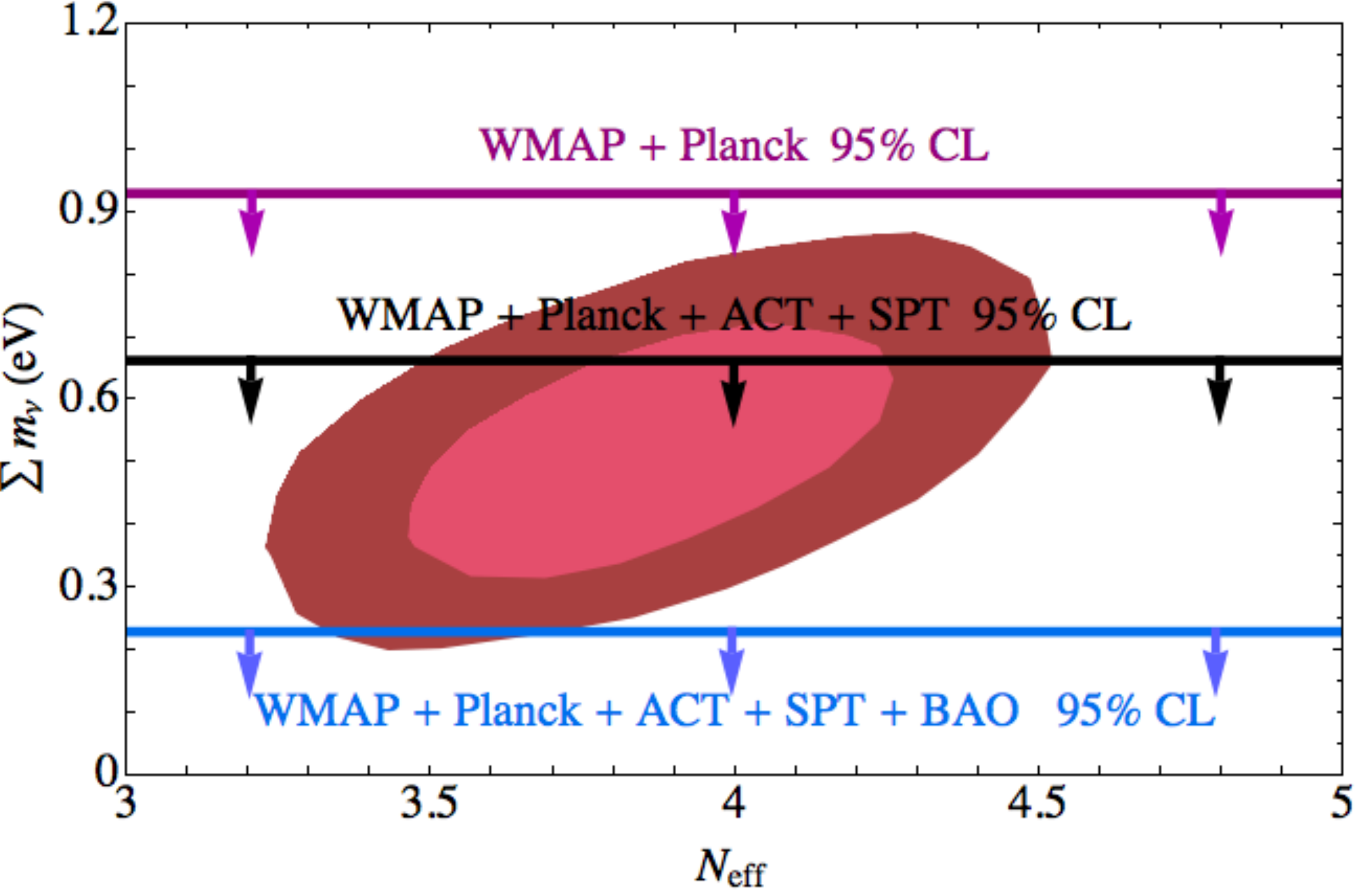}{0.99}
\end{minipage}
\begin{minipage}[t]{0.49\textwidth}
\postscript{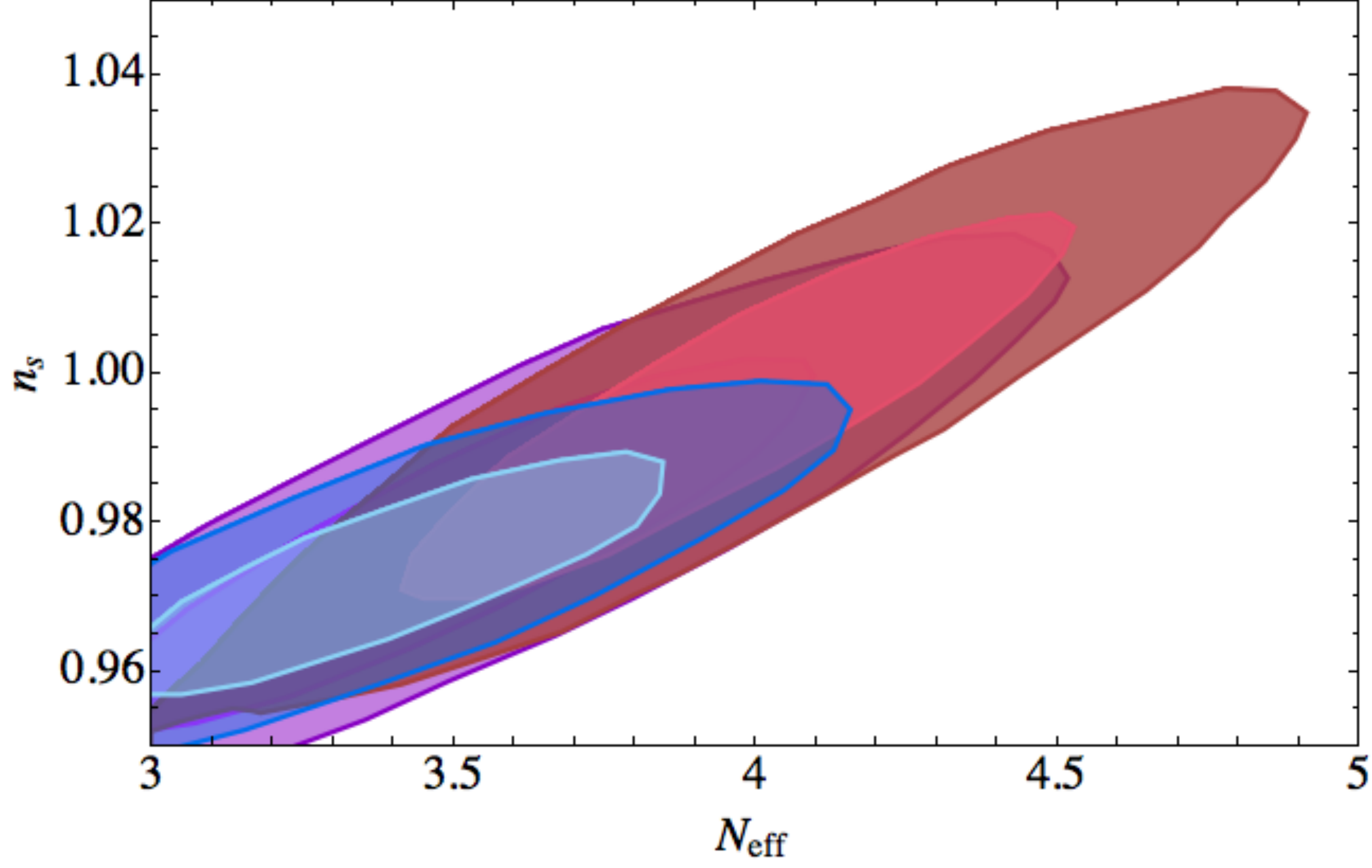}{0.99}
\end{minipage}
\hfill
\begin{minipage}[t]{0.49\textwidth}
\postscript{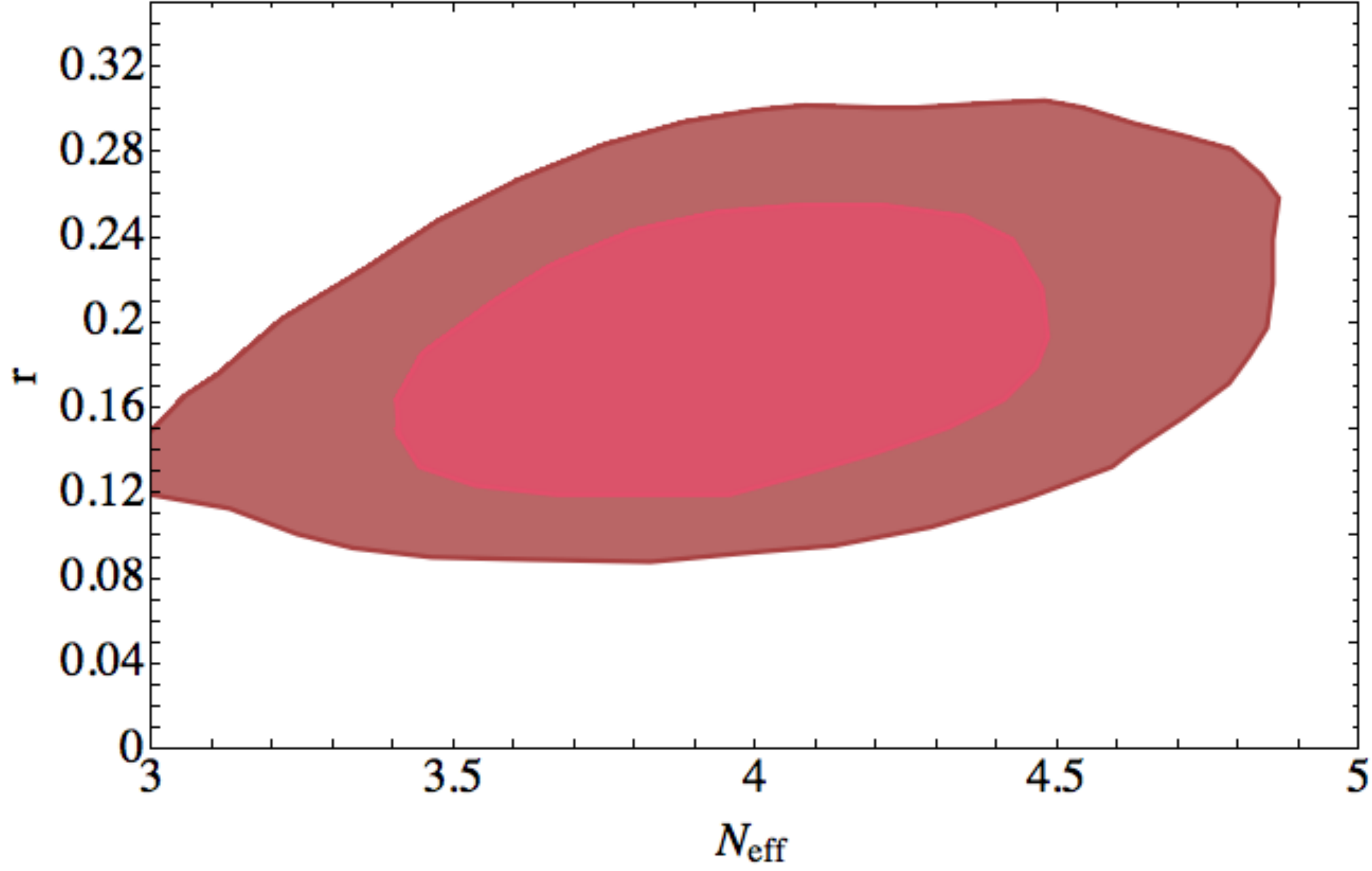}{0.99}
\end{minipage}
\caption{68\% and 95\%  confidence regions for the 9-parameter fit (burgundy). Also shown are the 68\% and 95\% confidence regions for $\Lambda$CDM + $N_{\rm eff}$, using  Planck + WMAP (purple), Planck + WMAP + BAO (blue) data. The horizontal lines indicate the 95\% CL upper limits on $\sum m_\nu$.}
\label{fig:dos} 
\end{figure}

Cosmological observations are sensitive to the total neutrino mass, that is to the sum of the three active neutrino masses
\begin{equation}
\Omega_\nu h^2 \, 94.1~{\rm eV} = (3.046/3)^{3/4} \, \sum m_\nu \, .
\end{equation}
Laboratory neutrino oscillation experiments indicate that at least two
species must be massive, and the sum of all three species must be
$\sum m_\nu > 0.055~{\rm eV}$~\cite{Beringer:1900zz}. Measurements
from the Planck temperature power spectrum in combination with
low-$\ell$ polarization measurements from WMAP 9-year
data~\cite{Bennett:2012zja} (WMAP + Planck) yield a 95\% CL upper
limit on the sum of the three active neutrino masses of $\sum m_\nu<
0.933~{\rm eV}$~\cite{Ade:2013zuv}. When data from the Atacama
Cosmology Telescope
(ACT)~\cite{Dunkley:2010ge,Dunkley:2013vu,Das:2013zf} and the South
Pole Telescope
(SPT)~\cite{Keisler:2011aw,Story:2012wx,Reichardt:2012yj} are
incorporated into the analysis the 95\% CL neutrino mass upper limit
is considerably improved, $\sum m_\nu< 0.663~{\rm eV}$~\cite{Ade:2013zuv}. Finally, with the addition of Baryon
Acoustic Oscillation (BAO) measurements from the Sloan Digital Sky
Survey (SDSS)-II Data Release
7~\cite{Percival:2009xn,Padmanabhan:2012hf}, from the WiggleZ
survey~\cite{Blake:2011en}, from the Baryon Acoustic Spectroscopic
Survey (BOSS)~\cite{Dawson:2012va} (one of the four surveys of
SDSS-III~\cite{Eisenstein:2011sa} Data Release
9~\cite{Anderson:2012sa}), and from 6dF Galaxy
Survey~\cite{Beutler:2011hx} the constraint on the neutrino mass is strongly tightened:
$\sum m_\nu< 0.230~{\rm eV}$ at 95\%CL~\cite{Ade:2013zuv}.

Before proceeding we note that the Planck CMB temperature anisotropy
spectrum is also in conflict with measurements of the local
Universe. Unexpectedly, the best multi-parameter fit of Planck data
yields a Hubble constant $h= 0.674 \pm 0.012$~\cite{Ade:2013zuv}, a
result which deviates by more than 2$\sigma$ from the value obtained
with the Hubble Space Telescope (HST), $h = 0.738 \pm
0.024$~\cite{Riess:2011yx}. (Herein we adopt the usual convention of
  writing the Hubble constant at the present day as $H_0 = 100 \
  h~{\rm km} \ {\rm s}^{-1} \ {\rm Mpc}^{-1}$.) The impact of the
Planck $h$ estimate is particularly important in the determination of
$N_{\rm eff}$. Combining observations of the CMB with data from BAO,
the Planck Collaboration reported $N_{\rm eff} = 3.30 \pm
0.27$~\cite{Ade:2013zuv}. However, if the value of $h$ is not allowed
to float in the fit, but instead is frozen to the value determined from the
maser-cepheid-supernovae distance ladder, $h = 0.738 \pm 0.024$, the Planck CMB data
then gives $N_{\rm eff} =3.62 \pm 0.25$, which suggests new
neutrino-like physics (at around the $2.3 \sigma$
level)~\cite{Ade:2013zuv}.

In Figs.~\ref{fig:uno} and \ref{fig:dos} we compare aftermath of the
multiparameter fit of Ref.~\cite{Dvorkin:2014lea}, generalized here
for extra massless r.d.o.f. ({\it i.e.,} $\{ \Omega_{\rm CDM} h^2,$
$\Omega_b h^2,$ $\tau, \, \Theta_{\rm s},\, A_s \,, n_s,\, r,\, N_{\rm
  eff},\, \sum m_\nu\})$, with the results reported by the
Planck~\cite{Ade:2013uln,Ade:2013zuv} and BICEP2~\cite{Ade:2014xna}
collaborations.  The data sets consider in the 9-parameter fit include various CMB
observations: temperature power spectrum from Planck~\cite{Ade:2013zuv}, WMAP 9-year
polarization~\cite{Bennett:2012zja}, ACT/SPT high multipole power
spectra~\cite{Das:2013zf,Keisler:2011aw,Reichardt:2012yj}, and the
BICEP2 BB and EE polarization band powers~\cite{Ade:2014xna}. In
addition a collection of local data sets has been included: the $H_0$
inference from the maser-cepheid-supernovae distance ladder and BAO
measurements~\cite{Padmanabhan:2012hf,Blake:2011en,Anderson:2012sa}. From
Fig.~\ref{fig:uno} we can see that for the 9-parameter fit, the 68\%
CL values of the parameter which controls the amplitude of the
gravitation wave signal are roughly equal to those reported by the
BICEP2 Collaboration~\cite{Ade:2014xna}. However, the allowed region
of the $r - n_s$ plane is shifted to larger values of the scalar spectral
index, and it is consistent with exact scale invariance ($n_s =1$)~\cite{Harrison:1969fb,Zeldovich:1972zz,Peebles:1970ag}. In
contrast to the results from Planck in the absence of running
($\alpha_s = 0$) the outcome of the multiparameter fit favors
inflationary models with $V''>0$. The allowed region of the $h - r$
parameter space is in agreement with the value of $h$ measured by both
Planck and HST~\cite{Ade:2013zuv}. In the upper
left panel of Fig.~\ref{fig:dos} we show the allowed 68\% and 95\%
confidence regions for $(h - N_{\rm eff})$. Unlike the previous results
reported by the Planck Collaboration on the basis of CMB data alone
(which are consistent with $N_{\rm eff} = 3.046$) the multiparameter
fit favors extra r.d.o.f. The 68\% CL contour yield $N_{\rm eff} =
3.86 \pm 0.25$. The associated 95\% CL region for the sum of the three
active neutrino masses saturates current limits (see upper right panel
of Fig.~\ref{fig:dos}.) Finally, the higher effective number of
r.d.o.f. is in agreement with the scalar spectral index
(Fig.~\ref{fig:dos} lower left) and the tensor-to-scalar ratio
(Fig.~\ref{fig:dos} lower right).

In closing, we note that the constraint on $N_{\rm eff}$ derived above
is in excellent agreement with the value inferred from BBN
observations: $N_{\rm eff} =
3.71^{+0.47}_{-0.45}$~\cite{Steigman:2012ve}. We take this agreement
as evidence favoring extra r.d.o.f. both during BBN and CMB epochs. In
the next section we interpret the observed excess, $\Delta N = 0.81
\pm 0.25$, in the context of the $E_6$ grand unified model, which is
largely well motivated by the  measured scale of inflation.

\section{Right-Handed Neutrinos with Milli-Weak Interactions}

In addition to the $2.984 \pm 0.009$ $\nu_ L$ species measured from
the width for invisible decays of the $Z$ boson~\cite{:2005ema} there
could also exist $\nu_R$ states that are sterile, {\it i.e.}  singlets
of the SM gauge group and therefore insensitive to weak
interactions. Such sterile states are predicted in models involving
additional TeV-scale $Z'$ gauge bosons, which allow for milli-weak
interactions of the $\nu_R$. If the $\nu_R$ carry a non-zero $U(1)$
charge, then the $U(1)$ symmetry forbids them from obtaining a
Majorana mass much larger than the $U(1)$-breaking scale. Therefore,
in most of these models there are no Majorana mass terms and the
$\nu_R$ states, which are almost massless, become the Dirac partners
of the SM $\nu_L$ species.

In this section we pursue a study to correlate the expected increase
in the universe expansion rate due to the presence of such light Dirac
neutrinos with ongoing searches of $Z'$ gauge bosons at the CERN's
LHC. A critical input for such an analysis is the relation between the
relativistic degrees of r.d.o.f.  and the temperature of the
primordial plasma. This relation is complicated because the
temperature which is of interest for right-handed neutrino decoupling
from the heat bath may lay in the vicinity of the quark-hadron
cross-over transition.  To connect the temperature to an effective
number of r.d.o.f. we make use of some high statistics lattice
simulations of a QCD plasma in the hot phase, especially the behavior
of the entropy during the changeover~\cite{Bazavov:2009zn}.

\subsection{Theoretical Considerations}

To develop our program in the simplest way, we
follow~\cite{Ellis:1985fp,GonzalezGarcia:1989py,Lopez:1989dh,Barger:2003zh,SolagurenBeascoa:2012cz}
and make use of extra $U(1)$ symmetries embedded in a grand unified
exceptional $E_6$ group, with breaking pattern
\begin{equation}
E_6 \to SO(10) \times
U(1)_\psi \to SU(5) \times U(1)_\psi \times U(1)_\chi \, .
\end{equation}
In $E_6$, each family of left-handed fermions is promoted to a fundamental $\bf 27$-plet, 
which decomposes under
$E_6\to SO(10)\to SU(5)$ as 
\be
{\bf 27}\to  {\bf 16}+{\bf 10}+{\bf 1}\to ({\bf 10}+{\bf 5^*}+{\bf 1})+({\bf 5}+{\bf 5^*})+{\bf 1},
\label{e6}
\ee as described in Table~\ref{tab:uno}~\cite{Hewett:1988xc,Langacker:2008yv}.  In addition to the SM
fermions, each $\bf 27$-plet contains two SM singlets, $\nu^c$ and
$S$, which may be charged under the extra $U(1)$ symmetries. The
$\nu^c$ can be interpreted as the conjugate of the right-handed
neutrino. There is also an exotic color-triplet quark $D$ and its conjugate $D^c$, both of which are $SU(2)$ singlets,
and a pair of color-singlet $SU(2)$-doublet exotics,
$H_u=\douba{H_u^+}{H_u^0}$ and $H_d=\douba{H_d^0}{H_d^-}$. All the
exotic fields are singlets or non-chiral under the SM,
but may be chiral under the extra $U(1)$ symmetries.

It is usually assumed that the gauge sector contains only one $U (1)$
symmetry at low energies. Thus, there is a continuum of possible
models where the new gauge boson couple to a linear combination of
$Q_\chi$ and $Q_\psi$ parametrized by a mixing angle
$\theta_{E_6}$. The resultant $U(1)'$ charge is then
\begin{equation}
Q_i = Q_\chi \cos \theta_{E_6} + Q_\psi \, \sin \theta_{E_6} \, .
\end{equation}
In this work we focus on the particular case $\theta_{E_6} = 0$, in 
which $S$ does not couple to the $Z'$. This model provides
a test basis for $Z'$ searches at the
ATLAS~\cite{Aad:2012hf,ATLAS:2012pu} and
CMS~\cite{Chatrchyan:2012it,Chatrchyan:2012oaa,Chatrchyan:2013qha}
experiments.

\begin{table}[ht]
\caption{Decomposition of the $E_6$ fundamental representation
of left-handed fermions
${\bf 27}$ under $SO(10)$ and $SU(5)$, and the $U(1)$ charges $Q_i$
for 
particular choices of $\theta_{E_6}$:  $U(1)_{\chi}$ with $\theta_{E_6} = 0$,
$U(1)_{\psi}$ with $\theta_{E_6} = \pi/2$, $U(1)_{\eta}$ with $\theta_{E_6} = \pi -
{\rm arctan} \sqrt{5/3} \approx 0.71 \, \pi$, inert $U(1)_I$ with $\theta_{E_6} = 
{\rm arctan} \sqrt{3/5} \approx 0.21 \, \pi$,
neutral-$N$ $U(1)_N$ with $\theta_{E_6} = 
{\rm arctan} \sqrt{15} \approx 0.42 \, \pi$, and  secluded sector
$U(1)_S$ with $\theta_{E_6} = 
{\rm arctan} \sqrt{15/9} \approx 0.13 \, \pi$~\cite{Robinett:1982tq,Langacker:1984dc,Witten:1985xc,Ma:1995xk,Erler:2002pr}.
\label{tab:uno}}
\begin{center}
\begin{tabular}{cccccccc}
\hline \hline 
~~$SO(10)$~~ & ~~$SU(5)$~~ & ~~$2 \sqrt{10} Q_{\chi}$~~ & ~~$2 \sqrt{6}
Q_{\psi}$~~ & ~~$2 \sqrt{15} Q_{\eta}$~~  & ~~$2Q_I$~~ & ~~$2
\sqrt{10} Q_N$~~ & ~~$2 \sqrt{15} Q_S$~~\\
\hline
16   &   $10~ (u,d,{u^c}, {e^+} )$ & $-$1 & $\phantom{-}$1  & $-$2&$\phantom{-}$0  &$\phantom{-}$1& $-{1/2}$\\
            &   ${5^\ast}~ ( d^c, \nu ,e^-)$  & $\phantom{-}$3  &$\phantom{-}$1 & $\phantom{-}$1&$-1$  & $\phantom{-}$2  & $\phantom{-}$4  \\
            &   $\nu^c$             & $-5$ &$\phantom{-}$1  & $-5$  &$\phantom{-}$1  &$\phantom{-}$0    & $-5$  \\
\hline
       10   &   $5~(D,H_u)$    &$\phantom{-}$2  & $-2$ & $\phantom{-}$4   &$\phantom{-}$0  & $-$2    & $\phantom{-}$1  \\
            &   ${5^\ast} ~(D^c, H_d)$ & $-2$ &$-2$ & $\phantom{-}$1 &$\phantom{-}$1& $-$3  & $-{7/2}$\\
\hline
       1    &   $1~ S$                  &  $\phantom{-}$0 & $\phantom{-}$4 & $-5$&$-1$ & $\phantom{-}$5 &$\phantom{-}5/2$\\
\hline
\hline
\end{tabular}
\end{center}
\end{table}

\subsection{Confronting Neutrino Cosmology with LHC Data}

In line with our stated plan, we now use the favored
value of $N_{\rm eff}$ to calculate the range of decoupling
temperature. We begin by first establishing the contribution of right-handed
neutrinos to $N_{\rm eff}$, that is $\Delta N_\nu$ as a function of
the $\nu_R$ decoupling temperature. Taking into account the isentropic
heating of the rest of the plasma between $T_{\nu_R}^{\rm dec}$ and
$T_{\nu_L}^{\rm dec}$ decoupling temperatures we obtain~\cite{Anchordoqui:2012qu}
\begin{equation}
 \Delta N_\nu = 3 \left( \frac{g_s(T_{\nu_L}^{\rm
  dec})}{g_s(T_{\nu_R}^{\rm dec})}\right) ^{4/3} \,, 
\end{equation}
where $g_s(T)$ is the
effective number of interacting (thermally coupled) r.d.o.f. at
temperature $T$; {\it e.g.}, $g_s(T_{\nu_L}^{\rm dec}) = 43/4$.  At
energies above the deconfinement transition towards the quark gluon
plasma, quarks and gluons are the relevant fields for the QCD sector,
such that the total number of SM r.d.o.f. is $g_s = 61.75$.  As the
universe cools down, the SM plasma transitions to a regime where
mesons and baryons are the pertinent degrees of freedom. Precisely,
the relevant hadrons present in this energy regime are pions and
charged kaons, such that $g_s = 19.25$~\cite{Brust:2013ova}. This
significant reduction in the degrees of freedom results from the rapid
annihilation or decay of any more massive hadrons which may have
formed during the transition. The quark-hadron crossover transition
therefore corresponds to a large redistribution of entropy into the
remaining degrees of freedom. Concretely, the effective number of
interacting r.d.o.f. in the plasma at temperature $T$ is given by 
\begin{equation}
g_s
(T) \simeq r (T) \left(N_B+ \frac{7}{8} N_{\rm F} \right),
\end{equation}
with $N_{\rm B} = 2$ for each real vector field and $N_{\rm F} = 2$
for each spin-$\frac{1}{2}$ Weyl field. The coefficient $r (T)$ is
unity for leptons, two for photon contributions, and is the ratio $s(T
)/s_{\rm SB}$ for the quark-gluon plasma. Here, $s(T)$ $(s_{\rm SB})$
is the actual (ideal Stefan-Boltzmann) entropy shown in
Fig~\ref{fig:dos}. For $150~{\rm MeV} < T < 500~{\rm MeV}$,  we
parametrize the entropy rise during the confinement-deconfinement
changeover by
\begin{equation}
\frac{s}{T^3}  \simeq  \frac{42.82}{\sqrt{392 \pi}} e^{-\frac{\left( T_{\rm MeV} - 151 \right)^2}{392}}  + \left( \frac{195.1}{T_{\rm MeV} - 134} \right)^2 \ 18.62 \frac{e^{195.1/(T_{\rm MeV} -134)}} {\left[
    e^{195.1/(T_{\rm MeV} -134)} - 1\right]^2} \, .
\label{soverT}
\end{equation}
For the same energy range, we obtain
\begin{equation}
g_s(T) \simeq 47.5 \ r(T) + 19.25 \, .
\label{gdet}
\end{equation}
In Fig.~\ref{fig:dos} we show $g_s(T)$ as given by (\ref{gdet}). Our
parametrization is in very good agreement with the phenomenological
estimate of~\cite{Laine:2006cp,Steigman:2012nb}.

\begin{figure}[!t]
\postscript{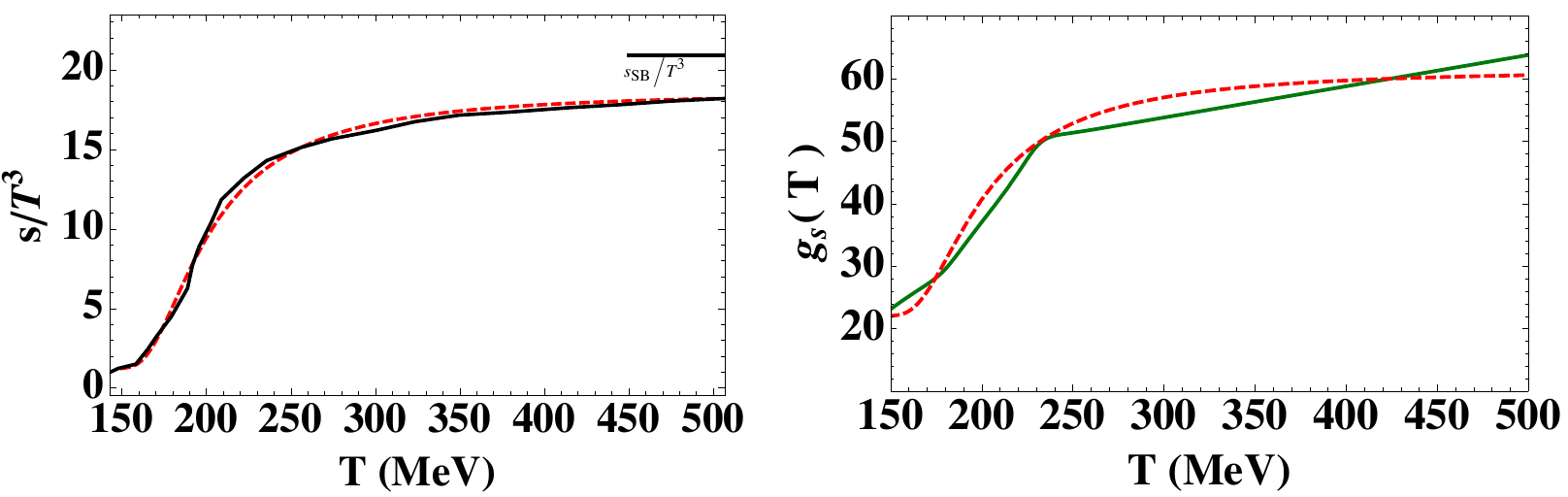}{0.99}
  \caption{{\bf Left.} The parametrization of the entropy density
    given in Eq.~(\ref{soverT})  (dashed line) superposed on the
    result from high statistics lattice
    simulations~\cite{Bazavov:2009zn} (solid line). {\bf Right.}~Comparison of $g_s (T)$ obtained using Eq.~(\ref{gdet}) (dashed
    line) and the phenomenological estimate of~\cite{Laine:2006cp,Steigman:2012nb} (solid line).}
\label{fig:tres}
\end{figure}

If relativistic particles are present that have decoupled from the
photons, it is necessary to distinguish between two kinds of r.d.o.f.:
those associated with the total energy density $g_\rho$, and those
associated with the total entropy density $g_s$. Since the quark-gluon
energy density in the plasma has a similar $T$ dependence to that of
the entropy (see Fig. 7 in~\cite{Bazavov:2009zn}), we take $ g_\rho
(T) \simeq r (T) \left(N_B+ \frac{7}{8} N_{\rm F} \right)$.

The right-handed neutrino decouples  from the plasma when its mean
free path becomes greater than the Hubble radius at that time 
\begin{equation}
 \Gamma(T_{\nu_R}^{\rm dec}) = H(T_{\nu_R}^{\rm dec}) \, ,
\label{haim1}
\end{equation}
where
\begin{equation}
\Gamma (T_{\nu_R}^{\rm dec}) = {\cal K} \ \frac{1}{8} \ \left(\frac{\overline  g}{M_{Z'}}
\right)^4 \ (T_{\nu_R}^{\rm dec})^5 \ \sum_{i=1}^6 {\cal N}_i \,,
\label{Gamma}
\end{equation}
is the $\nu_R$ interaction rate, 
\begin{eqnarray}
H(T^{\rm dec}_{\nu_R}) & = & 1.66 \sqrt{g_\rho} \ (T_{\nu_R}^{\rm dec})^2
/M_{\rm Pl} \nonumber \\
& \simeq &
1.66 \sqrt{g_s (T^{\rm dec}_{\nu_L}) } \
\frac{(T^{\rm dec}_{\nu_R})^2}{M_{\rm
    Pl}} \ \left( \frac{3}{\Delta
  N_\nu} \right)^{3/8} \,,
\label{Hubble}
\end{eqnarray}
is the Hubble expansion rate at the $\nu_R$ decoupling temperature
\begin{equation}
\overline g \equiv 
\left(\frac{\sum_{i =1}^6  {\cal N}_i g_i^2 g_6^2}{\sum_{i =i}^6  {\cal
    N}_i} \right)^{1/4}\, ,
\label{gbarra}
\end{equation}
${\cal N}_i$ is the number of chiral states, $g_i = g_0 \, Q_i$ are
the chiral couplings of the $Z'$ for the 6 relevant species (see
below), and the constant ${\cal K} = 0.5 \ (2.5)$ for annihilation
(annihilation + scattering)~\cite {Anchordoqui:2011nh}. In the second
line of (\ref{Hubble}) we set $g_s \simeq g_\rho$. In conformity with
grand unification we follow~\cite{Barger:2003zh} and choose
\begin{equation}
g_{0} = \sqrt{\frac{5}{3}} \ g_2  \ \tan \theta_W \sim 0.46 \,,
\end{equation}
with $g_2$ the $SU(2)_L$ coupling. (Note that for the $N$ model, the effective coupling
$\overline g$ has a  similar stength.)

\begin{figure}[tbp]
\begin{minipage}[t]{0.49\textwidth}
\postscript{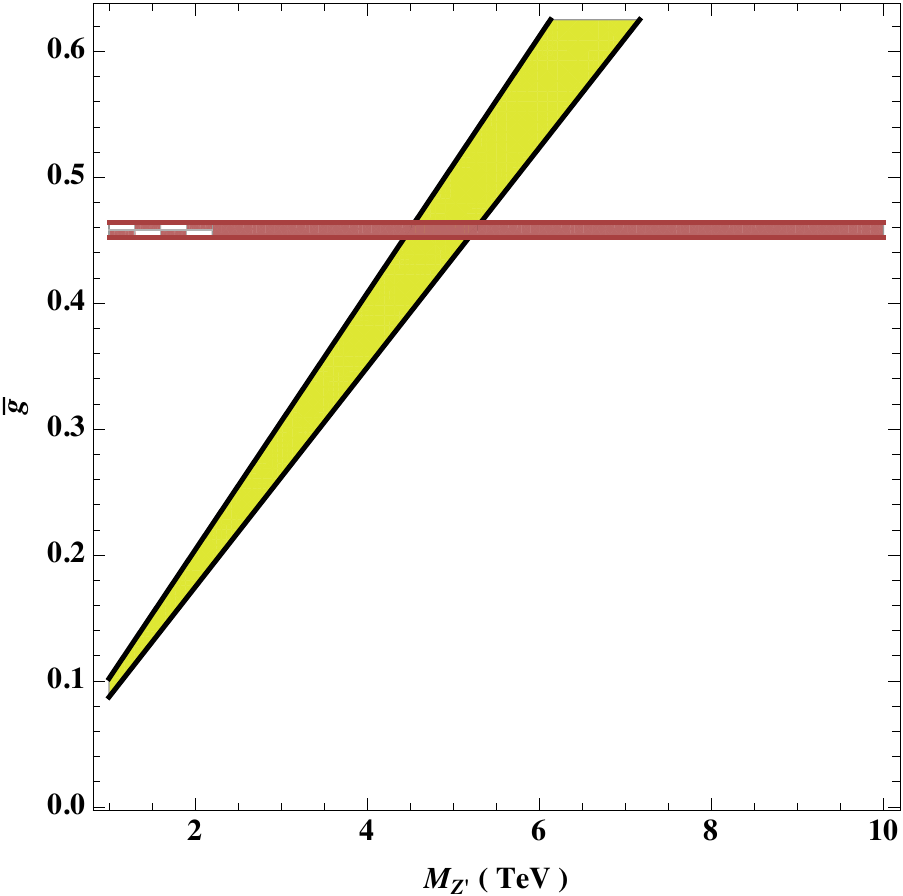}{0.99}
\end{minipage}
\hfill
\begin{minipage}[t]{0.49\textwidth}
\postscript{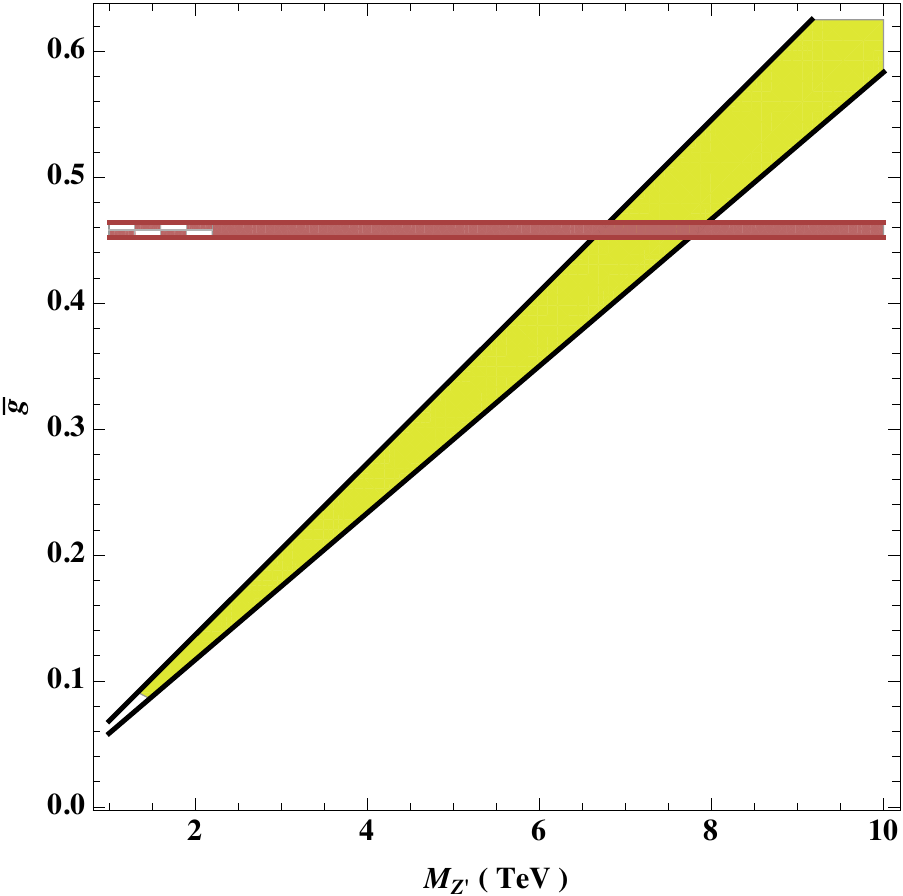}{0.99}
\end{minipage}
\caption{The (yellow) shaded area shows the $68\%$ confidence region
  allowed from decoupling requirements to accommodate $\Delta N = 0.81
  \pm 0.25$,  The horizontal line dictates the effective coupling $\overline g$
  for $Z'_\chi$; the cross-hatched part of the line  reflects the LHC experimental limits on the mass of the gauge
  boson. We have taken ${\cal K} =0.5$ (left) and ${\cal K} = 2.5$
  (right).}
\label{fig:cuatro}
\end{figure}

The physics of interest takes place in the quark gluon plasma itself
so that we will restrict ourselves to the following fermionic fields
in the visible sector, $ \left[ 3 u_R \right] + \left[ 3 d_R \right] +
\left[ 3s_R\right ] + \left[ 3 \nu_L + e_L + \mu_L \right] + \left[
  e_R + \mu_R \right]+ \left[ 3 u_L + 3d_L + 3s_L\right]+ \left[ 3
  \nu_R \right]$, and their contribution to $g_\rho$. This amounts to
28 Weyl fields, translating to 56 fermionic r.d.o.f.; $\sum_{i=1}^6 
{\cal N}_i = 28$. 

Substituting (\ref{Gamma}) and
(\ref{Hubble}) into (\ref{haim1}) we obtain
\begin{displaymath}
\frac{\overline g}{M_{Z'}} = \left( \frac{3}{\Delta N_\nu}\right)^{3/32}
\left(\frac{13.28 \ \sqrt{g_s(T_{\nu_L}^{\rm dec})}}{M_{\rm Pl} \ {\cal
      K} \ (T_{\nu_R}^{\rm dec})^3 \  \sum_{i=1}^6 {\cal N}_i} \right)^{1/4} 
\end{displaymath}
and
\begin{equation}
 \Delta N_\nu 
 =  \left[\frac{5.39 \times 10^{-6}}{{\cal K} \sum_{i=1}^6 \mathcal{N}_i} \left(  \frac{M_{Z'}}{{\rm TeV}} \ \frac{1}{\bar{g}} \right)^4 \ \left(\frac{{\rm GeV}}{T_{\nu_R}}\right)^3 \right]^{8/3}  \, .
\label{hitichi}
\end{equation}
In Fig.~\ref{fig:cuatro} we show the region of the parameter space
allowed from decoupling requirements to accommodate contributions of
$\Delta N_\nu = 0.81 \pm 0.25$. This region is in agreement with LHC experimental limits
on $M_{Z'}$ for null signals for enhancements in dilepton or dijet 
searches~\cite{Aad:2012hf,ATLAS:2012pu,Chatrchyan:2012it,Chatrchyan:2012oaa,Chatrchyan:2013qha}. The model also provides a testable prediction, $\overline
g \simeq 0.46$ and $4.5~{\rm TeV} \alt M_{Z'} \alt 7.5~{\rm TeV}$, within
the LHC14 discovery reach.

\section{Conclusions}

Aside from exhibiting temperature fluctuations of one part in
$10^{5}$, the CMB is partially polarized.  The parity-odd
polarization, or B-mode, arises from primordial tensor fluctuations,
which manifest as gravitational waves. The temperature map provided by
the Planck mission constrains the value of tensor-to-scalar
perturbations, $r<0.11$ at 95 CL. However, recent BICEP2 B-mode
polarization data is in tension with this constrain, as imply a value
$r=0.20^{+0.07}_{-0.05}$.  Very recently, it was suggested that
additional relativistic degrees of freedom beyond the three active
neutrinos and photons can help to relieve this tension: the data favor
an effective number of light neutrino species $N_{\rm eff} = 3.86 \pm
0.25$.

We have shown that we can accommodate the required $N_{\rm eff}$ with
the contribution from the right-handed partners of the three,
left-handed, SM neutrinos (living in the fundamental representation of
$E_6$). The six additional fermionic r.d.o.f. can be suppressed to
levels in compliance with the favored $N_{\rm eff}$, because the
milli-weak interactions of these Dirac states (through their coupling
to a TeV-scale $Z'$ gauge boson) may allow the $\nu_R$'s to decouple
much earlier, at a higher temperature, than their left-handed
counterparts.  If the $\nu_R$'s decouple during the quark-hadron
crossover transition, they are considerably cooler than the $\nu_L$'s
and contribute less than 3 extra ``equivalent neutrinos'' to the early
Universe energy density. For decoupling in this transition region, the
3\,$\nu_R$ generate $\Delta N_{\nu} = 3(T_{\nu_R}^{\rm
  dec}/T_{\nu_L}^{\rm dec})^{4} < 3$, extra relativistic degrees of
freedom.  These requirements strongly constrain the mass of the heavy
vector field.  Consistency (within $1\sigma$) with $N_{\rm eff}$ is achieved for an effective coupling $\overline g =
0.46$ and $Z'$ mass in the range $4.5~{\rm TeV} < M_{Z'} < 7.5~{\rm
  TeV}$. The model is fully predictive and can be confronted with
future data from LHC14.

\section*{Acknowledgments}
This work was supported in part by the US NSF grants: CAREER
PHY1053663 (LAA) and PHY-0757959 (HG); NASA Grant No. NNX13AH52G
(LAA); and the UWM RGI (BJV).


\begin{thebibliography}{99}

\bibitem{Beringer:1900zz} 
  J.~Beringer {\it et al.}  [Particle Data Group Collaboration],
  Phys.\ Rev.\ D {\bf 86}, 010001 (2012).

\bibitem{Guth:1980zm} 
  A.~H.~Guth,
  Phys.\ Rev.\ D {\bf 23}, 347 (1981).

\bibitem{Linde:1981mu} 
  A.~D.~Linde,
  Phys.\ Lett.\ B {\bf 108}, 389 (1982).


\bibitem{Linde:1983gd} 
  A.~D.~Linde,
  Phys.\ Lett.\ B {\bf 129}, 177 (1983).


\bibitem{Albrecht:1982wi} 
  A.~Albrecht and P.~J.~Steinhardt,
  Phys.\ Rev.\ Lett.\  {\bf 48}, 1220 (1982).


\bibitem{Freese:1990rb} 
  K.~Freese, J.~A.~Frieman and A.~V.~Olinto,
  Phys.\ Rev.\ Lett.\  {\bf 65}, 3233 (1990).



\bibitem{Mukhanov:1981xt} 
  V.~F.~Mukhanov and G.~V.~Chibisov,
  JETP Lett.\  {\bf 33}, 532 (1981)
  [Pisma Zh.\ Eksp.\ Teor.\ Fiz.\  {\bf 33}, 549 (1981)].

\bibitem{Hawking:1982cz}
  S.~W.~Hawking,
  Phys.\ Lett.\ B {\bf 115}, 295 (1982).

\bibitem{Bardeen:1983qw} 
  J.~M.~Bardeen, P.~J.~Steinhardt and M.~S.~Turner,
  Phys.\ Rev.\ D {\bf 28}, 679 (1983).



\bibitem{Leach:2002ar} 
  S.~M.~Leach, A.~R.~Liddle, J.~Martin and D.~J Schwarz,
  Phys.\ Rev.\ D {\bf 66}, 023515 (2002)
  [astro-ph/0202094].



\bibitem{Ade:2014xna} 
  P.~A.~R.~Ade {\it et al.}  [BICEP2 Collaboration],
  arXiv:1403.3985 [astro-ph.CO].


\bibitem{Ade:2013uln} 
  P.~A.~R.~Ade {\it et al.}  [Planck Collaboration],
  arXiv:1303.5082 [astro-ph.CO].

\bibitem{Hinshaw:2012aka} 
  G.~Hinshaw {\it et al.}  [WMAP Collaboration],
  Astrophys.\ J.\ Suppl.\  {\bf 208}, 19 (2013)
  [arXiv:1212.5226 [astro-ph.CO]].



\bibitem{Ashoorioon:2014nta} 
  A.~Ashoorioon, K.~Dimopoulos, M.~M.~Sheikh-Jabbari and G.~Shiu,
  arXiv:1403.6099 [hep-th].

\bibitem{Ko:2014bka} 
  P.~Ko and Y.~Tang,
  arXiv:1404.0236 [hep-ph].

\bibitem{Smith:2014kka} 
  K.~M.~Smith, C.~Dvorkin, L.~Boyle, N.~Turok, M.~Halpern, G.~Hinshaw and B.~Gold,
  arXiv:1404.0373 [astro-ph.CO].


\bibitem{Giusarma:2014zza} 
  E.~Giusarma, E.~Di Valentino, M.~Lattanzi, A.~Melchiorri and O.~Mena,
  arXiv:1403.4852 [astro-ph.CO].


\bibitem{Zhang:2014dxk} 
  J.~-F.~Zhang, Y.~-H.~Li and X.~Zhang,
  arXiv:1403.7028 [astro-ph.CO].


\bibitem{Dvorkin:2014lea} 
  C.~Dvorkin, M.~Wyman, D.~H.~Rudd and W.~Hu,
  arXiv:1403.8049 [astro-ph.CO].




\bibitem{Starobinsky:1985ww} 
  A.~A.~Starobinsky,
  Sov.\ Astron.\ Lett.\  {\bf 11}, 133 (1985).



\bibitem{Anchordoqui:2014uua} See {\it e.g.},
  L.~A.~Anchordoqui, V.~Barger, H.~Goldberg, X.~Huang and D.~Marfatia,
  arXiv:1403.4578 [hep-ph].




\bibitem{Steigman:1977kc} 
  G.~Steigman, D.~N.~Schramm and J.~E.~Gunn,
  Phys.\ Lett.\ B {\bf 66}, 202 (1977).



\bibitem{Mangano:2005cc} 
  G.~Mangano, G.~Miele, S.~Pastor, T.~Pinto, O.~Pisanti and P.~D.~Serpico,
  Nucl.\ Phys.\ B {\bf 729}, 221 (2005)
  [hep-ph/0506164].

\bibitem{Bennett:2012zja} 
  C.~L.~Bennett {\it et al.}  [WMAP Collaboration],
  Astrophys.\ J.\ Suppl.\  {\bf 208}, 20 (2013)
  [arXiv:1212.5225 [astro-ph.CO]].


\bibitem{Ade:2013zuv} 
  P.~A.~R.~Ade {\it et al.}  [Planck Collaboration],
  arXiv:1303.5076 [astro-ph.CO].




\bibitem{Dunkley:2010ge} 
  J.~Dunkley, R.~Hlozek, J.~Sievers, V.~Acquaviva, P.~A.~R.~Ade, P.~Aguirre, M.~Amiri and J.~W.~Appel {\it et al.},
  Astrophys.\ J.\  {\bf 739}, 52 (2011)
  [arXiv:1009.0866 [astro-ph.CO]].


\bibitem{Dunkley:2013vu} 
  J.~Dunkley, E.~Calabrese, J.~Sievers, G.~E.~Addison, N.~Battaglia, E.~S.~Battistelli, J.~R.~Bond and S.~Das {\it et al.},
  arXiv:1301.0776 [astro-ph.CO].


\bibitem{Das:2013zf} 
  S.~Das, T.~Louis, M.~R.~Nolta, G.~E.~Addison, E.~S.~Battistelli, J R.~Bond, E.~Calabrese and D.~C.~M.~J.~Devlin {\it et al.},
  arXiv:1301.1037 [astro-ph.CO].

\bibitem{Keisler:2011aw} 
  R.~Keisler, C.~L.~Reichardt, K.~A.~Aird, B.~A.~Benson, L.~E.~Bleem, J.~E.~Carlstrom, C.~L.~Chang and H.~M.~Cho {\it et al.},
  Astrophys.\ J.\  {\bf 743}, 28 (2011)
  [arXiv:1105.3182 [astro-ph.CO]].


\bibitem{Reichardt:2012yj}
  C.~L.~Reichardt, B.~Stalder, L.~E.~Bleem, T.~E.~Montroy, K.~A.~Aird, K.~Andersson, R.~Armstrong and M.~L.~N.~Ashby {\it et al.},
  Astrophys.\ J.\  {\bf 763} (2013) 127
   [Astrophys.\ J.\  {\bf 763} (2013) 127]
  [arXiv:1203.5775 [astro-ph.CO]].

\bibitem{Story:2012wx} 
  K.~T.~Story, C.~L.~Reichardt, Z.~Hou, R.~Keisler, K.~A.~Aird, B.~A.~Benson, L.~E.~Bleem and J.~E.~Carlstrom {\it et al.},
  Astrophys.\ J.\  {\bf 779}, 86 (2013)
  [arXiv:1210.7231 [astro-ph.CO]].







\bibitem{Percival:2009xn} 
  W.~J.~Percival {\it et al.}  [SDSS Collaboration],
  Mon.\ Not.\ Roy.\ Astron.\ Soc.\  {\bf 401}, 2148 (2010)
  [arXiv:0907.1660 [astro-ph.CO]].


\bibitem{Padmanabhan:2012hf} 
  N.~Padmanabhan, X.~Xu, D.~J.~Eisenstein, R.~Scalzo, A.~J.~Cuesta, K.~T.~Mehta and E.~Kazin,
  Mon.\ Not.\ Roy.\ Astron.\ Soc.\  {\bf 427}, no. 3, 2132 (2012)
  [arXiv:1202.0090 [astro-ph.CO]].


\bibitem{Blake:2011en} 
  C.~Blake, E.~Kazin, F.~Beutler, T.~Davis, D.~Parkinson, S.~Brough, M.~Colless and C.~Contreras {\it et al.},
  Mon.\ Not.\ Roy.\ Astron.\ Soc.\  {\bf 418}, 1707 (2011)
  [arXiv:1108.2635 [astro-ph.CO]].

\bibitem{Dawson:2012va} 
  K.~S.~Dawson {\it et al.}  [BOSS Collaboration],
  Astron.\ J.\  {\bf 145}, 10 (2013)
  [arXiv:1208.0022 [astro-ph.CO]].

\bibitem{Eisenstein:2011sa} 
  D.~J.~Eisenstein {\it et al.}  [SDSS Collaboration],
  Astron.\ J.\  {\bf 142}, 72 (2011)
  [arXiv:1101.1529 [astro-ph.IM]].


\bibitem{Anderson:2012sa} 
  L.~Anderson, E.~Aubourg, S.~Bailey, D.~Bizyaev, M.~Blanton, A.~S.~Bolton, J.~Brinkmann and J.~R.~Brownstein {\it et al.},
  Mon.\ Not.\ Roy.\ Astron.\ Soc.\  {\bf 427}, no. 4, 3435 (2013)
  [arXiv:1203.6594 [astro-ph.CO]].


\bibitem{Beutler:2011hx} 
  F.~Beutler, C.~Blake, M.~Colless, D.~H.~Jones, L.~Staveley-Smith, L.~Campbell, Q.~Parker and W.~Saunders {\it et al.},
  Mon.\ Not.\ Roy.\ Astron.\ Soc.\  {\bf 416}, 3017 (2011)
  [arXiv:1106.3366 [astro-ph.CO]].


\bibitem{Riess:2011yx} 
  A.~G.~Riess {\it et al.},
  Astrophys.\ J.\  {\bf 730}, 119 (2011)  [Erratum ibid.\  {\bf 732}, 129 (2011)]
  [arXiv:1103.2976 [astro-ph.CO]].


\bibitem{Harrison:1969fb} 
  E.~R.~Harrison,
  Phys.\ Rev.\ D {\bf 1}, 2726 (1970).


\bibitem{Zeldovich:1972zz} 
  Y.~.B.~Zeldovich,
  Mon.\ Not.\ Roy.\ Astron.\ Soc.\  {\bf 160}, 1P (1972).


\bibitem{Peebles:1970ag} 
  P.~J.~E.~Peebles and J.~T.~Yu,
  Astrophys.\ J.\  {\bf 162}, 815 (1970).


\bibitem{Steigman:2012ve} 
  G.~Steigman,
  Adv.\ High Energy Phys.\  {\bf 2012}, 268321 (2012)
  [arXiv:1208.0032 [hep-ph]].





\bibitem{:2005ema}
 S.~Schael {\it et al.}  [ALEPH Collaboration, DELPHI Collaboration, L3 Collaboration, OPAL Collaboration, SLD Collaboration, LEP Electroweak Working Group, and SLD Electroweak and Heavy Flavour Groups],
  Phys.\ Rept.\  {\bf 427}, 257 (2006)
  [arXiv:hep-ex/0509008].



\bibitem{Bazavov:2009zn} 
  A.~Bazavov  {\it et al.},
  Phys.\ Rev.\ D {\bf 80}, 014504 (2009)
  [arXiv:0903.4379 [hep-lat]].




\bibitem{Ellis:1985fp} 
  J.~R.~Ellis, K.~Enqvist, D.~V.~Nanopoulos and S.~Sarkar,
  Phys.\ Lett.\ B {\bf 167}, 457 (1986).

\bibitem{GonzalezGarcia:1989py}
  M.~C.~Gonzalez-Garcia and J.~W.~F.~Valle,
  Phys.\ Lett.\ B {\bf 240}, 163 (1990).

\bibitem{Lopez:1989dh} 
  J.~L.~Lopez and D.~V.~Nanopoulos,
  Phys.\ Lett.\ B {\bf 241}, 392 (1990).

\bibitem{Barger:2003zh}
  V.~Barger, P.~Langacker and H.~S.~Lee,
  Phys.\ Rev.\  D {\bf 67}, 075009 (2003)
  [arXiv:hep-ph/0302066].

\bibitem{SolagurenBeascoa:2012cz} 
  A.~Solaguren-Beascoa and M.~C.~Gonzalez-Garcia,
  Phys.\ Lett.\ B {\bf 719}, 121 (2013)
  [arXiv:1210.6350 [hep-ph]].



\bibitem{Robinett:1982tq} 
  R.~W.~Robinett and J.~L.~Rosner,
  Phys.\ Rev.\ D {\bf 26}, 2396 (1982).



\bibitem{Langacker:1984dc} 
  P.~Langacker, R.~W.~Robinett and J.~L.~Rosner,
  Phys.\ Rev.\ D {\bf 30}, 1470 (1984).

\bibitem{Witten:1985xc} 
  E.~Witten,
  Nucl.\ Phys.\ B {\bf 258}, 75 (1985).



\bibitem{Ma:1995xk} 
  E.~Ma,
  Phys.\ Lett.\ B {\bf 380}, 286 (1996)
  [hep-ph/9507348].

\bibitem{Erler:2002pr} 
  J.~Erler, P.~Langacker and T.~-j.~Li,
  Phys.\ Rev.\ D {\bf 66}, 015002 (2002)
  [hep-ph/0205001].

\bibitem{Hewett:1988xc} 
  J.~L.~Hewett and T.~G.~Rizzo,
  Phys.\ Rept.\  {\bf 183}, 193 (1989).


\bibitem{Langacker:2008yv} 
  P.~Langacker,
  Rev.\ Mod.\ Phys.\  {\bf 81}, 1199 (2009)
  [arXiv:0801.1345 [hep-ph]].




\bibitem{Aad:2012hf}  
  G.~Aad {\it et al.}  [ATLAS Collaboration],
  JHEP {\bf 1211}, 138 (2012)
  [arXiv:1209.2535 [hep-ex]].

\bibitem{ATLAS:2012pu} 
  G. Aad {\it et al.}  [ATLAS Collaboration],
  JHEP {\bf 1301}, 029 (2013).




\bibitem{Chatrchyan:2012it}
  S.~Chatrchyan {\it et al.}  [CMS Collaboration],
  Phys.\ Lett.\ B {\bf 714}, 158 (2012)
  [arXiv:1206.1849 [hep-ex]].



\bibitem{Chatrchyan:2012oaa} 
  S.~Chatrchyan {\it et al.}  [CMS Collaboration],
  Phys.\ Lett.\ B {\bf 720}, 63 (2013)
  [arXiv:1212.6175 [hep-ex]].

\bibitem{Chatrchyan:2013qha} 
  S.~Chatrchyan {\it et al.}  [CMS Collaboration],
  Phys.\ Rev.\ D {\bf 87}, no. 11, 114015 (2013)
  [arXiv:1302.4794 [hep-ex]].







\bibitem{Anchordoqui:2012qu} 
  L.~A.~Anchordoqui, H.~Goldberg and G.~Steigman,
  Phys.\ Lett.\ B {\bf 718}, 1162 (2013)
  [arXiv:1211.0186 [hep-ph]].







\bibitem{Brust:2013ova} 
  C. Brust, D.E. Kaplan and M.T. Walters,
  arXiv:1303.5379. 





\bibitem{Laine:2006cp} 
  M. Laine and Y. Schroder,
  Phys.\ Rev.\ D {\bf 73}, 085009 (2006)
  [hep-ph/0603048].

\bibitem{Steigman:2012nb}
  G. Steigman, B. Dasgupta and J.F. Beacom,
  Phys.\ Rev.\ D {\bf 86}, 023506 (2012)
  [arXiv:1204.3622 [hep-ph]].


\bibitem{Anchordoqui:2011nh} 
  L.~A.~Anchordoqui and H.~Goldberg,
  Phys.\ Rev.\ Lett.\  {\bf 108}, 081805 (2012)
  [arXiv:1111.7264 [hep-ph]].




\end{thebibliography}
\end{document}